\documentclass[twocolumn]{aastex63}

\usepackage[caption=false]{subfig}
\usepackage{mathtools,amssymb}
\newcommand{\vsh}{v_{\rm sh}}

\newcommand{\w}[1]{v_{\rm A,#1}}

\newcommand{\rsh}{R_{\rm sh}}

\submitjournal{ApJ}

\shorttitle{RS Ophiuchi}

\begin{document}

\title{Evidence for multiple shocks from the $\gamma$-ray emission of RS Ophiuchi}

\correspondingauthor{Rebecca Diesing}
\email{rrdiesing@uchicago.edu}

\author[0000-0002-6679-0012]{Rebecca Diesing}
\affiliation{Department of Astronomy and Astrophysics, The University of Chicago, 5640 S Ellis Ave, Chicago, IL 60637, USA}

\author[0000-0002-4670-7509]{Brian D.~Metzger}
\affil{Department of Physics and Columbia Astrophysics Laboratory, Columbia University, New York, NY 10027, USA}
\affil{Center for Computational Astrophysics, Flatiron Institute, 162 5th Ave, New York, NY 10010, USA} 

\author[0000-0001-8525-3442]{Elias Aydi}
\affiliation{Center for Data Intensive and Time Domain Astronomy, Department of Physics and Astronomy, Michigan State University, East Lansing, MI 48824, USA}

\author[0000-0002-8400-3705]{Laura Chomiuk}
\affiliation{Center for Data Intensive and Time Domain Astronomy, Department of Physics and Astronomy, Michigan State University, East Lansing, MI 48824, USA}

\author[0000-0003-1336-4746]{Indrek Vurm}
\affiliation{Tartu Observatory, University of Tartu, Observatooriumi 1, 61602 T{$\bar o$}ravere, Estonia}

\author[0000-0002-1030-8012]{Siddhartha Gupta}
\affiliation{Department of Astronomy and Astrophysics, The University of Chicago, 5640 S Ellis Ave, Chicago, IL 60637, USA}

\author[0000-0003-0939-8775]{Damiano Caprioli}
\affiliation{Department of Astronomy and Astrophysics, The University of Chicago, 5640 S Ellis Ave, Chicago, IL 60637, USA}
\affiliation{Enrico Fermi Institute, The University of Chicago, Chicago, IL 60637, USA}

\begin{abstract}

 In August of 2021, Fermi-LAT, H.E.S.S., and MAGIC detected GeV and TeV $\gamma$-ray emission from an outburst of recurrent nova RS Ophiuchi. This detection represents the first very high energy $\gamma$-rays observed from a nova, and opens a new window to study particle acceleration. 
 Both H.E.S.S. and MAGIC described the observed $\gamma$-rays as arising from a single, external shock. 
 In this paper, we perform detailed, multi-zone modeling of RS Ophiuchi's 2021 outburst including a self-consistent prescription for particle acceleration and magnetic field amplification. 
 We demonstrate that, contrary to previous work, a single shock cannot simultaneously explain RS Ophiuchi's GeV and TeV emission, particularly the spectral shape and distinct light curve peaks.   Instead, we put forward a model involving multiple shocks that reproduces the observed $\gamma$-ray spectrum and temporal evolution.  
 The simultaneous appearance of multiple distinct velocity components in the nova optical spectrum over the first several days of the outburst supports the presence of distinct shocks, which may arise either from the strong latitudinal dependence of the density of the external circumbinary medium (e.g., in the binary equatorial plane versus the poles) or due to internal collisions within the white dwarf ejecta (as powers the $\gamma$-ray emission in classical novae). \\ 

\end{abstract}

\section{Introduction} \label{sec:intro}

A major discovery by the {\it Fermi} Large Area Telescope (LAT) was that novae$-$multi-wavelength transients produced by  non-terminal thermonuclear explosions on the surface of white dwarfs accreting hydrogen-rich material from a donor star$-$are sources of luminous $\sim$GeV $\gamma$-ray emission (e.g., \citealt{abdo+10,Ackermann+14,Cheung+16,Franckowiak+18}).  Novae come in two varieties: ``classical'' novae, when the donor is a main-sequence or moderately-evolved star overflowing its Roche Lobe onto the white dwarf; and ``embedded'' or ``symbiotic'' novae, when the donor is instead a giant star with a dense wind.

The first nova detected in $\gamma$-rays, V407 Cyg, was of the embedded type, with a Mira-like red giant donor \citep{abdo+10}.  The $\gamma$-rays from this event, which started around the time of optical maximum and lasted about two weeks, were interpreted by \citet{abdo+10} as non-thermal emission from relativistic particles (ions or electrons) accelerated to high energies at the shock wave generated as the nova ejecta collided with the dense circumbinary environment of the giant companion (e.g., \citealt{Chomiuk+12}).  \citet{martin+13} modeled the diffusive acceleration of particles from the resulting shock wave, finding that inverse Compton emission by relativistic electrons interacting with the red giant optical light are the dominant $\gamma$-ray emission source; they also found that a ``density enhancement'' in the binary plane was needed to match the data in addition to the standard red giant wind (\citealt{booth+16} describe a possible origin for such an equatorial density enhancement generated by the white dwarf's accretion of the giant wind).

Given the relative rarity of novae with giant companions, $\gamma$-ray detections of novae were predicted to be rare (see the discussion in \citealt{abdo+10}).  However, this expectation was upended when {\it Fermi} LAT began to detect additional Galactic novae, even of the more common classical variety (\citealt{Ackermann+14,Cheung+16,Franckowiak+18}; a total of $\gtrsim 15$ LAT-detected classical novae to date).  Given the lack of a dense wind from a main-sequence star, the shocks in classical novae are likely ``internal'', i.e., occurring between distinct components of nova ejecta.  Indeed, multi-wavelength observations from X-ray (e.g., \citealt{Mukai&Ishida01,Takei+09,Nelson+19,Gordon+21}) to radio (e.g., \citealt{chomiuk+14,Weston+16}), show internal shocks to be common if not ubiquitous in classical novae (see \citealt{Chomiuk+21} for a recent review). 
Resolved radio imaging (e.g., \citealt{chomiuk+14}) supports a picture in which the shock interaction is caused by a quasi-spherical high-velocity outflow or wind from the white dwarf which collides with a slower outflow concentrated in the plane of the binary and released earlier in the eruption.  Additional evidence for this interpretation comes from time-resolved optical spectra, which show evidence for multiple velocity components which exist simultaneously, the faster of which first appears around optical maximum and near the onset of the $\gamma$-ray emission (e.g., \citealt{Aydi+20b, hachisu+22}).  

The gas surrounding the internal shocks in classical novae is sufficiently dense to act as a ``calorimeter'' for converting their thermal UV/X-ray emission (\citealt{metzger+14}) and non-thermal accelerated particles (\citealt{Vurm&Metzger18,Martin+18}) into reprocessed optical and $\gamma$-ray emission, respectively.  This should result in a contribution to the nova optical light curve from shocks that tracks the $\gamma$-ray light curve \citep{Metzger+15}, consistent with that observed in the few novae for which such a measurement has been possible \citep{Li+17,Aydi+20a}. 

RS Ophiuchi (RS Oph) is a binary consisting of a red giant orbiting a white dwarf with a 454 day period \citep{Brandi+09}, which has undergone a nova eruption every $\sim 10-20$ years for over the last century \citep{Schaefer10}. Its 2006 eruption occurred before the launch of {\it Fermi} LAT, but signatures of shock interaction were observed in impressive detail at X-ray, infrared, and radio wavelengths \citep{Sokoloski+06, Bode+06, OBrien+06, Das+06, Tatischeff07, Evans+07, Rupen+08}. 
Its most recent eruption, beginning on 2021 Aug 8, qualified RS Ophiuchi as the second embedded nova detected at high-significance by {\it Fermi} LAT (\citealt{cheung+22}), and also as the first nova detected at $\sim$ TeV energies, by the Atmospheric Cherenkov Telescopes H.E.S.S. \citep{HESS22} and MAGIC \citep{MAGIC22}.  No classical nova has yet been detected at TeV energies, though relatively few upper limits have been reported in the literature (e.g., by MAGIC and HAWC; \citealt{Ahnen+15,Albert+22}) and there are theoretical reasons to believe that the high neutral gas densities surrounding the shock may limit the maximum particle energy \citep{reville+07,metzger+16}.    

Interestingly, the TeV light curve peak in RS Ophiuchi is delayed by several days relative to the peak of the GeV light curve, with the \emph{Fermi} LAT light curve peaking on 2021 Aug 9--10 \citep{cheung+22}  and the HESS light curve peaking on 2021 Aug 12 (\citealt{HESS22}; see their Figure 2). The HESS collaboration interprets the entire GeV-TeV emission in terms of hadronic particle acceleration and emission at a single shock, attributing the observed temporal delay between the peaks at different energies to the finite timescale required to accelerate ions to $\gtrsim$ TeV energies.
However, in this paper, we show that the $\gamma$-rays observed during RS Ophiuchi's 2021 outburst cannot be produced by a single, spherically symmetric shock. 
Instead, we put forward a scenario involving multiple shocks.  These shocks may be generated as the result of distinct velocity components of the nova ejecta interacting with the aspherical external environment.
This scenario can reproduce key features of the observed $\gamma$-rays without any ad-hoc modifications to the shape or maximum energy of the accelerated particle spectrum.

In Section \ref{sec:method} we describe our model for shock evolution, particle acceleration, and magnetic field amplification. 
We also introduce constraints on our model from optical spectroscopy.
We apply our model to RS Ophiuchi in Section \ref{sec:results1} and demonstrate how a single, spherical shock cannot reproduce the $\gamma$-ray observations. 
We show the results of our best-fit multi-shock model in Section \ref{sec:results2}, and discuss physical scenarios that might correspond to this model in Section \ref{sec:conclusion}.

Throughout this work, we assume a distance to RS Ophiuchi of 1.4 kpc \citep{barry+08}, consistent with  that used in \cite{HESS22}.

\section{Method} \label{sec:method}

To calculate RS Ophiuchi's expected $\gamma$-ray emission we use a multi-zone model of particle acceleration and photon production. 
A detailed description of this model can be found below.

\subsection{Shock Hydrodynamics} \label{subsec:hydro}

To estimate shock evolution, we use a self-similar formalism similar to that described in \cite{diesing+18}.
Namely, we assume that both the material ejected by the nova and the material swept up during expansion are confined to a thin shell behind the shock \cite[see, e.g.,][for examples of this \emph{thin-shell approximation}]{bisnovatyi-kogan+95, ostriker+88, bandiera+04}. The evolution of the shock is thus set by the density profile of the ambient medium, chosen to reproduce the $\gamma$-ray observations and to be consistent with the presence of a RG wind (see Section \ref{sec:results1}).

More specifically, we model RS Ophiuchi during two stages of evolution: the \emph{ejecta-dominated stage}, in which the mass of swept-up material is less than the ejecta mass and the shock expands freely, and the \emph{Sedov stage}, in which the swept-up mass exceeds the ejecta mass and the nova expands adiabatically. 
Energy is conserved throughout both stages such that, given a nova with energy $E_{\rm kin}$,
\begin{equation}
    v_{\rm sh} = \bigg(\frac{2E_{\rm kin}}{M_{\rm ej}+M_{\rm SU}}\bigg)^{1/2}.
\end{equation}
Here, $v_{\rm sh}$ is the forward shock velocity, $M_{\rm ej}$ is the ejecta mass, and $M_{\rm SU}$ is the swept up mass, given by,
\begin{equation}
    M_{\rm SU} = \int_{R_{\rm min}}^{R_{\rm sh}} 4\pi r^2\rho_0(r)dr.
\end{equation}

The above formalism applies to a single shock.  However, as we shall discuss in later section, the $\gamma$-ray emission and optical spectra of RS Ophiuchi both suggest the presence of multiple shock components.  These shocks may arise due to distinct mass ejection events from the white dwarf (as in the case of classical novae; see discussion in Sec.~\ref{sec:intro}), and/or a single ejecta shell interacting with non-spherically symmetric external medium.  In the latter case, we will in effect be applying the above formalism separately to distinct angular sectors (e.g., the polar versus equatorial region) over which we assume the density is approximately uniform.  

The above model assumes a stationary external medium, which is a good approximation for the slowly-expanding RG wind or circumbinary disk expected around RS Ophiuchi (e.g., \citealt{booth+16}).  However, it would less well approximate the dynamics of internal shocks between distinct ejecta components from the white dwarf (e.g., \citealt{metzger+14}), as are believed to power the GeV $\gamma$-ray emission from classical novae (Sec.~\ref{sec:intro}).  Nevertheless, the above framework may still provide a rough description of the internal shock case, provided that the initial shock velocity is interpreted as the relative velocity between the ejecta components.  

It is also worth noting that nova outbursts may also be described by a continuous wind as opposed to an instantaneous energy injection \citep[e.g., ][]{kato+22}. In this case, the hydrodynamic model used in our work does not apply and, importantly, a reverse shock may contribute to the observed $\gamma$-ray luminosity. However, the fact that the optical emission begins to decay before the GeV peak suggests that such extended energy injection is unlikely \citep[e.g., ][]{metzger+14, Li+17}. More specifically, after its maximum, the optical luminosity decays as $t^{-1.3}$, implying that a substantial portion of the optical energy is released near its peak. By the time of the GeV maximum, we do not expect significant ongoing energy injection from a wind, meaning that our instantaneous ejection model provides a good approximation for the hydrodynamics of the system.

Admittedly, it may still be possible for a reverse shock to contribute to the GeV emission at early times. However, such a contribution does not change the main conclusion of our work, that multiple shocks--be they forward shocks or forward and reverse shocks--are required to reproduce the observed gamma-ray emission.

\subsection{Particle Acceleration}

We model particle acceleration using a semi-analytic model of nonlinear diffusive shock acceleration that self-consistently accounts for magnetic field amplification and the dynamical back-reaction of accelerated particles on the shock \citep[see ][and references therein, in particular \cite{malkov97,malkov+00,blasi02,blasi04,amato+05, amato+06}]{caprioli+09a,caprioli+10b, caprioli12, diesing+19, diesing+21}. 
We assume that protons with momenta above $p_{\rm inj} \equiv \xi_{\rm inj}p_{\rm th}$ are injected into the acceleration process, where $p_{\rm th}$ is the thermal momentum and we choose $\xi_{\rm inj}$ to produce CR pressure fractions $\sim 10\%$ (though we note that the non-thermal acceleration efficiency in classical novae is typically measured to be closer to $\sim 1\%$; e.g., \citealt{Metzger+15,Aydi+20a}).  We also calculate the proton maximum energy self-consistently by requiring that the diffusion length (assuming Bohm diffusion) of the most energetic particles accelerated be equal to 5$\%$ of the shock radius.

This model produces an instantaneous distribution of protons accelerated at each timestep of nova evolution. 
Instantaneous electron distributions are calculated from these proton distributions following the analytical approximation in \cite{zirakashvili+07}. 
We then shift and weight each instantaneous distribution to account for adiabatic and, in the case of electrons, synchrotron losses \citep[see][for more details]{caprioli+10a,morlino+12,diesing+19}. We then sum these weighted distributions to yield the cumulative, multi-zone spectrum of non-thermal particles accelerated by our model novae.

We also account for proton-proton losses by calculating the collision rate for each instantaneous distribution (i.e., each expanding shell of protons) at each timestep, assuming the collisional cross-section parameterized in \cite{kafexhiu+14} and a target proton density equal to the adiabatically expanded post-shock density of that shell. 
We further assume that a proton loses half its energy in a single collision (i.e., we assume an inelasticity $\kappa = 0.5$, consistent with \citealt{martin+13}), and modify each instantaneous proton distribution accordingly.

Note that, unless the ambient density is quite large, we do not expect significant proton-proton losses on timescales of a few days\footnote{This is distinct from in classical novae, where the higher densities associated with internal shocks and greater ejecta masses often satisfy the calorimeric constraint (e.g., \citealt{Metzger+15,Li+17,Aydi+20a}).} since, for an interaction cross section of $\sim 30$ mb and $\kappa = 0.5$, the proton-proton loss time can be approximated as,
\begin{equation}
    t_{\rm pp} \simeq 260 \bigg(\frac{10^8 \rm{ cm}^{-3}}{n_0}\bigg) \text{day,}
\end{equation}
where $n_0$ is the number density of the ambient medium in front of the shock.   

\subsection{Photon Production}

To estimate photon spectra from our cumulative proton and electron distributions, we use the radiative processes code \emph{naima} \citep[][]{naima}.
\emph{Naima} computes the emission due to synchrotron, Bremsstrahlung, inverse Compton (IC) and neutral pion decay processes assuming arbitrary proton and electron distributions, as well as our chosen density profile(s). 
While the IC luminosity depends also on the radiation field chosen, we find that leptonic emission is subdominant regardless of our assumptions (see Section \ref{sec:results1}).

The main sources of opacity for GeV - TeV photons are pair production---on soft radiation fields in the RG wind and on nuclei in the nova ejecta.  The opacity throughout the nova ejecta (Bethe-Heitler process) for hydrogen-rich material is given by \citep[e.g.][]{Zdziarski&Svensson89}
\begin{equation}
\sigma_{\rm BH} = \frac{3}{8\pi}\alpha_{\rm fs}\sigma_{\rm T}\left[\frac{28}{9}{\rm ln}\left(2x\right) - \frac{218}{27}\right],
\end{equation}
where $x \equiv E_{\gamma}/m_e c^{2}$, $\alpha_{\rm fs} \simeq 1/137$ and $\sigma_{\rm T} \simeq 6.6\times 10^{-25}$ cm$^{-2}$ is the Thomson cross section.  Thus, the Bethe-Heitler optical depth through the RG wind,
\begin{align}
\tau_{\rm BH}(>r) &\simeq \int n_0\sigma_{\rm BH}dr \nonumber \\
&\simeq 1.9\times 10^{-5}\left(\frac{r}{1{\rm AU}}\right)^{-1}\left[\frac{28}{9}{\rm ln}\left(2x\right) - \frac{218}{27}\right],
\end{align}
is negligible.  The Bethe-Heitler optical depth is also likely to be irrelevant even if the density is enhanced by a factor $\sim 10^{3}$ on small scales $\lesssim$ AU.

$\gamma$-ray photons can also be attenuated due to $\gamma-\gamma$ pair production with ambient IR/optical/UV photons, again either from the RG or from the nova outburst itself. 

We calculate the optical depth at the location of the nova shock ($r_{\rm sh}$) for $\gamma$-rays due to absorption on the IR/optical radiation field  as,
\begin{equation}
\tau_{\gamma\gamma}(\epsilon_{\gamma}) = r_{\rm sh} \int_{(m_{\rm e} c^2)^2/\epsilon_{\gamma}} \sigma_{\gamma\gamma}(\epsilon_{\gamma},\epsilon_{\rm t})\frac{dn_{\rm opt}}{d\epsilon_{\rm t}} \, d\epsilon_{\rm t},
\label{eq:taugg}
\end{equation}
where $n_{\rm opt}$ is the number density of target photons with energy $\epsilon_{\rm t}$ and $\sigma_{\gamma\gamma}$ is the interaction cross section. The target radiation field is normalized to the bolometric optical luminosity, $L_{\rm opt}$ given by \cite{cheung+22} (their figure 9),
\begin{equation}
\frac{L_{\rm opt}}{4\pi r_{\rm sh}^{2}c} = \int \frac{dn_{\rm opt}}{d\epsilon_{\rm t}} \, \epsilon_{\rm t} \, d\epsilon_{\rm t}.
\end{equation}
The target radiation field is assumed to have a blackbody shape with $T_{\rm BB} = 10^4$~K. 
Our lack of knowledge of the actual broadband spectral shape is unlikely to introduce a significant error, since the primary targets for $\gamma$-rays of interest are within or close to the optical band (e.g., at 1~eV for 500~GeV $\gamma$-rays), hence the $\gamma-\gamma$ opacity is less sensitive to the precise shape of the target spectrum outside the observed range than it would be e.g. for $\epsilon_{\gamma} \gg 1$~TeV photons.

\begin{figure}[ht]
    \centering
    \includegraphics[width=0.48\textwidth, clip=true,trim= 10 20 40 35]{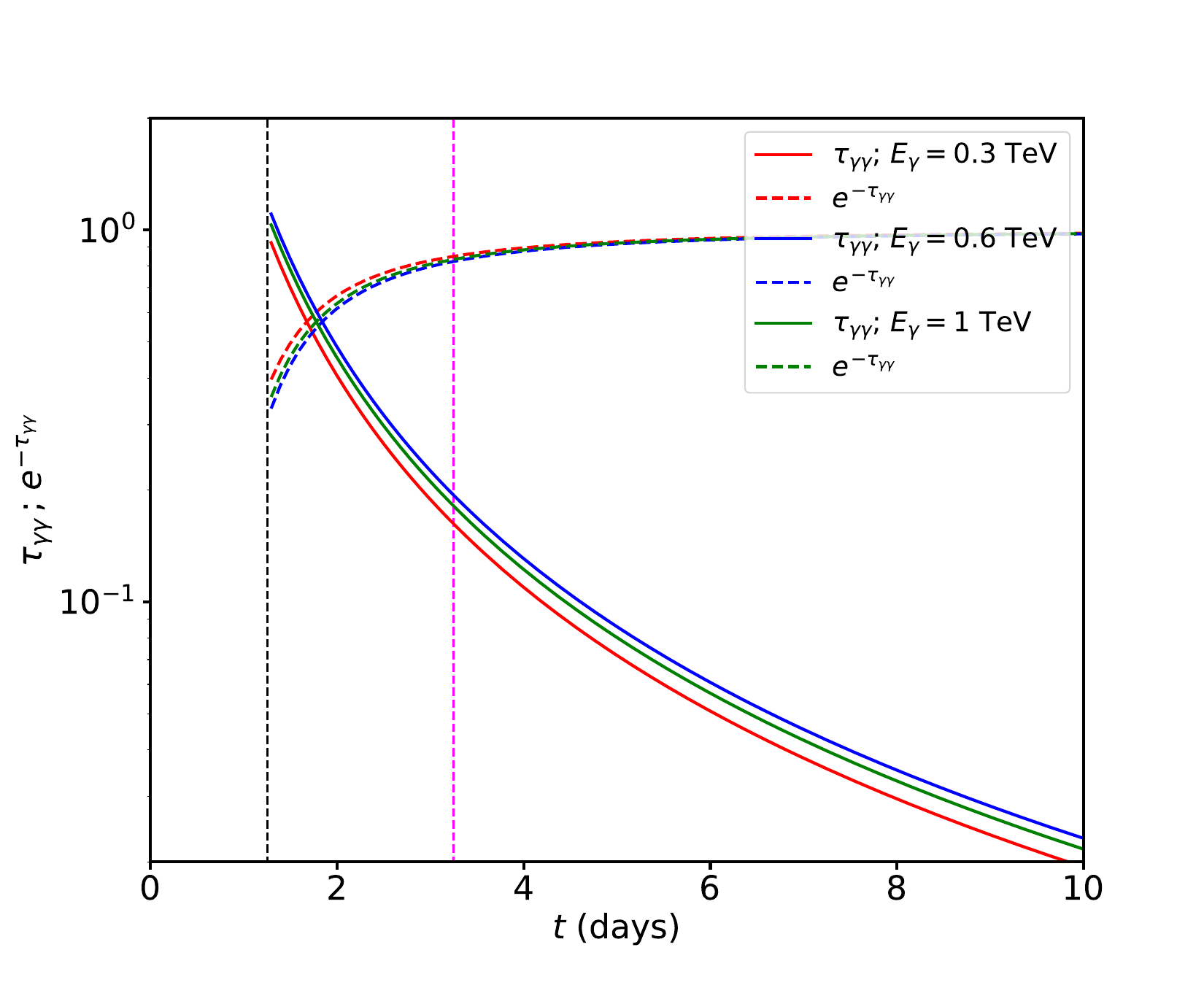}
    \caption{Optical depth (solid lines) and attenuation factor (dashed lines) for VHE $\gamma$-rays in the best-fit single-shock model (Table \ref{tab:singleshock}). The target radiation is assumed to have blackbody shape of $T_{\rm BB} = 10^4$~K and a bolometric luminosity given by \cite{cheung+22}. The vertical black and magenta dashed lines indicate the time of Fermi-LAT and H.E.S.S. luminosity peaks, respectively.}
    \label{fig:taugg}
\end{figure}

We show the results of this calculation in Figure \ref{fig:taugg}. $r_{\rm sh}$ is taken to be that of our best-fit model involving a single, external shock, as described in Section \ref{sec:results1}. While we do expect modest attenuation of TeV $\gamma$-rays (by a factor of $\sim 2$) at $t \simeq 1$ day, this attenuation is not sufficient to account for the order of magnitude rise in the TeV luminosity observed between days 1 and 4. Furthermore, since the $\gamma-\gamma$ opacity is negligible at the radius corresponding to the TeV luminosity peak, we neglect absorption in our emission estimates.

It is also worth noting that eq. (\ref{eq:taugg}) employs the angle-averaged $\gamma-\gamma$ cross-section, or equivalently, an isotropic target radiation field. If the optical radiation is produced at smaller radii than the VHE $\gamma$-rays, its beaming around the radial direction tends to suppress the attenuation of radially-directed $\gamma$-rays. In the limit of perfectly collimated targets, an escape cone exists for the $\gamma$-rays for any target radiation density; the (angle-averaged) attenuation is no longer exponential and a fraction of $\gamma$-rays can escape even if the isotropically-averaged $\tau_{\gamma\gamma}$ would indicate essentially complete suppression. 

\subsection{Magnetic Field Amplification} \label{subsec:mfa}

The streaming of energetic particles ahead of the shock is expected to excite various instabilities \citep[]{skilling75a,bell04,amato+09,bykov+13}, which amplify magnetic fields and enhance CR diffusion \citep{caprioli+14a,caprioli+14b,caprioli+14c}. 
This amplification has been inferred observationally in supernova remnants \citep[e.g.,][]{parizot+06, bamba+05, morlino+10, ressler+14}, and is expected to proceed in a similar manner in nova shocks. 

We model magnetic field amplification as in \cite{diesing+21} by assuming saturation of both the resonant streaming instability \citep[e.g.,][]{kulsrud+68,zweibel79,skilling75a, bell78a, lagage+83a}, and the non-resonant hybrid instability \citep{bell04}.
This prescription reproduces the magnetic fields inferred from X-ray observations of young supernova remnants \citep{volk+05,caprioli+08}. 

\cite{bell04} derives the saturation point of the non-resonant instability to be
\begin{equation}
    P_{\rm B1,Bell} = \frac{\vsh}{2c}\frac{P_{\rm CR}}{\gamma_{\rm CR} - 1}.
\end{equation}
Here, $\gamma_{\rm CR}=4/3$ is the CR adiabatic index. 
This saturation has been validated with hybrid simulations in \cite{zacharegkas+21,zacharegkas+22}. 

For fast shocks ($\gtrsim 100$ km s$^{-1}$ for typical nova parameters), the non-resonant instability dominates amplification \citep[see][ for a detailed discussion]{diesing+21}, so we assume that the total magnetic pressure immediately upstream of the shock is then $P_{\rm B,1} \simeq P_{\rm B1,res}$; 
moving further upstream, this pressure is taken to scale with the local $P_{\rm CR}(x)$.
Thus, for a fast shock capable of accelerating TeV particles, we expect the magnetic field in front of the shock to scale as $\vsh^{3/2}$, assuming the CR pressure scales with the ram pressure, $\rho_0 \vsh^2$.
Note that, given the strong magnetic field amplification in our model, the ambient magnetic field---a relatively unknown quantity---has a negligible impact on our results.

The analysis above assumes fully ionized gas surrounding the shock.  However, this may not be justified for sufficiently dense upstream gas, because the recombination timescale may be longer than the ionization timescale by UV/X-ray radiation from the shock (indeed, absorption and reprocessing of the shock thermal emission is key to powering nova optical emission from the shocks; \citealt{metzger+14}).  Upstream gas with a substantial neutral component can suppress the growth rate of the non-resonant instability \citep{bell04,reville+07}, thus reducing the maximum energy of the ions accelerated at nova shocks \citep{metzger+16}.

\subsection{Constraints from Optical Spectroscopy}

\label{sec:spectroscopy}

We now consider constraints on the properties of the ejecta from RS Ophiuchi based on optical spectroscopy.

We make use of publicly available high-resolution spectroscopy from the Astronomical Ring for Access to Spectroscopy (ARAS; \citealt{Teyssier_2019}) database, covering the first month of the eruption, starting from 0.5 days after $t_0$ (HJD 2459435.0745). We present the evolution of H$\beta$ and Fe II 5169 $\mathrm{\AA}$ line profiles during the first 30 days in Figures~\ref{Fig:Hbeta_complete},~\ref{Fig:Hbeta_abs},~\ref{Fig:FeII_complete}, and~\ref{Fig:FeII_abs}. 
The line shows multiple absorption components during the first week of the eruption. We identify at least 3 components: (1) an initial component that is present in the first spectrum taken 0.5 days after eruption discovery, and is at a velocity $v_1 = -2700$\,km\,s$^{-1}$; (2) a faster component characterized with a velocity $v_2 = -3700$\,km\,s$^{-1}$ that appears a day later and co-exists with the initial component for a few days. This component is only resolved in the Balmer lines; (3) a third component characterized by a velocity $v_3 = -1900$\,km\,s$^{-1}$ and appearing around days 3\,--\,4 in the Balmer lines, but which is prominent earlier in the Fe II lines (possibly as early as days 1\,--\,2). All these components show variations in their velocity/strength as the eruption evolves. Since these components co-exist in the same lines, they possibly originate in distinct regions of the nova ejecta or CSM. 

\begin{figure}
\begin{center}
  \includegraphics[width=\columnwidth]{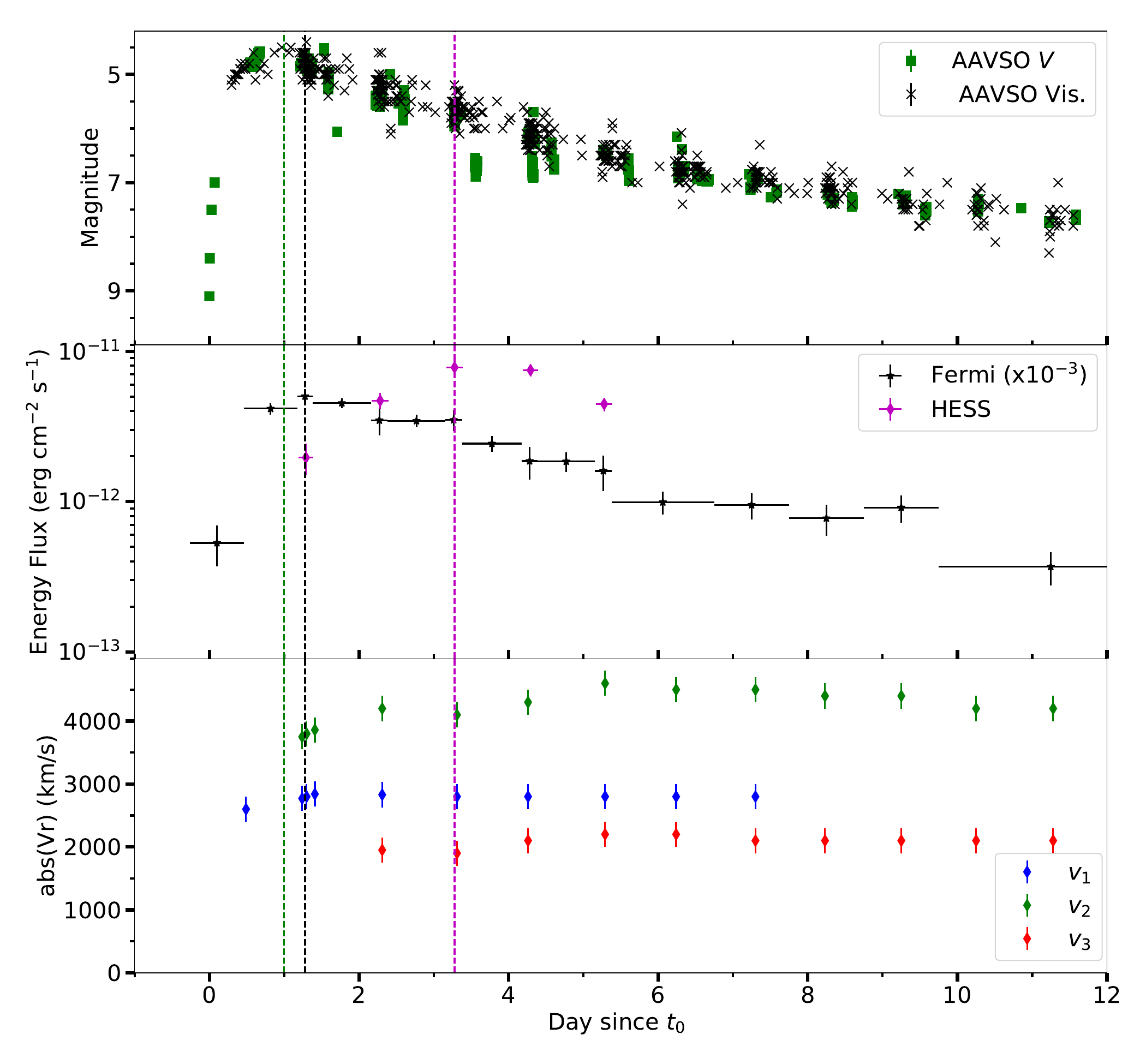}
\caption{\textit{Top}: the optical light curve of RS~Oph during the first 12 days of the eruption. \textit{Middle}: the high-energy Fermi and HESS light curves of RS~Oph during the first 12 days of the eruption. \textit{Bottom}: the velocity evolution of the different spectral components (labeled $v_1$, $v_2$, and $v_3$) observed in the Balmer and Fe II lines. The vertical green, black, and magenta dashed lines represent the time of optical, \textit{Fermi}, and HESS peaks, respectively.} 
\label{Fig:velocity_evolution}
\end{center}
\end{figure}

Figure~\ref{Fig:velocity_evolution} shows the evolution of these components relative to the optical and high-energy light curves.  One interpretation of these data is that the 2700\,km\,s$^{-1}$ component originates in a comparatively ``slow'' initial mass ejection phase concentrated in the binary equatorial plane, while the 3700\,km\,s$^{-1}$ component$-$which accelerates to $\gtrsim 4000$ km\,s$^{-1}$ over a few days$-$originates in a faster, radiation driven, wind that potentially expands more freely in the polar direction, in a scenario similar to that suggested for the early spectral evolution of classical novae (e.g., \citealt{Aydi+20b}). These components have also been reported in \citet{Molaro_etal_2022} and were observed in different species.  The slowest component with a velocity of 1900\,km\,s$^{-1}$ could then arise from swept-up shell formed by the nova ejecta interacting with the slowly expanding ($\approx 50$\,km\,s$^{-1}$) circumbinary material.

Another possible interpretation of these velocity components is elucidated in \cite{hachisu+22}. In this case, the components correspond to three different regions surrounding a single shock. However, this interpretation implies a shock velocity of 1900 km s${^-1}$, which is insufficient to accelerate TeV particles.

In what follows, we use the velocities of these ejecta components to motivate those assumed in our shock models. 
To keep matters simple, we will only consider systems with a maximum of two velocity components. 
However, we choose shock velocities that are reasonably consistent with those inferred from optical data. 
Our main objective here is to demonstrate that the optical observations of distinct ejecta components motivate the presence of multiple shocks which, as we shall argue, is needed to interpret the $\gamma$-ray data.  

\section{The Single-Shock Scenario} \label{sec:results1}

\begin{figure}[ht]
  \centering
  \subfloat[\label{fig:evo}]{%
      \includegraphics[width=0.48\textwidth, clip=true,trim= 10 30 40 35]{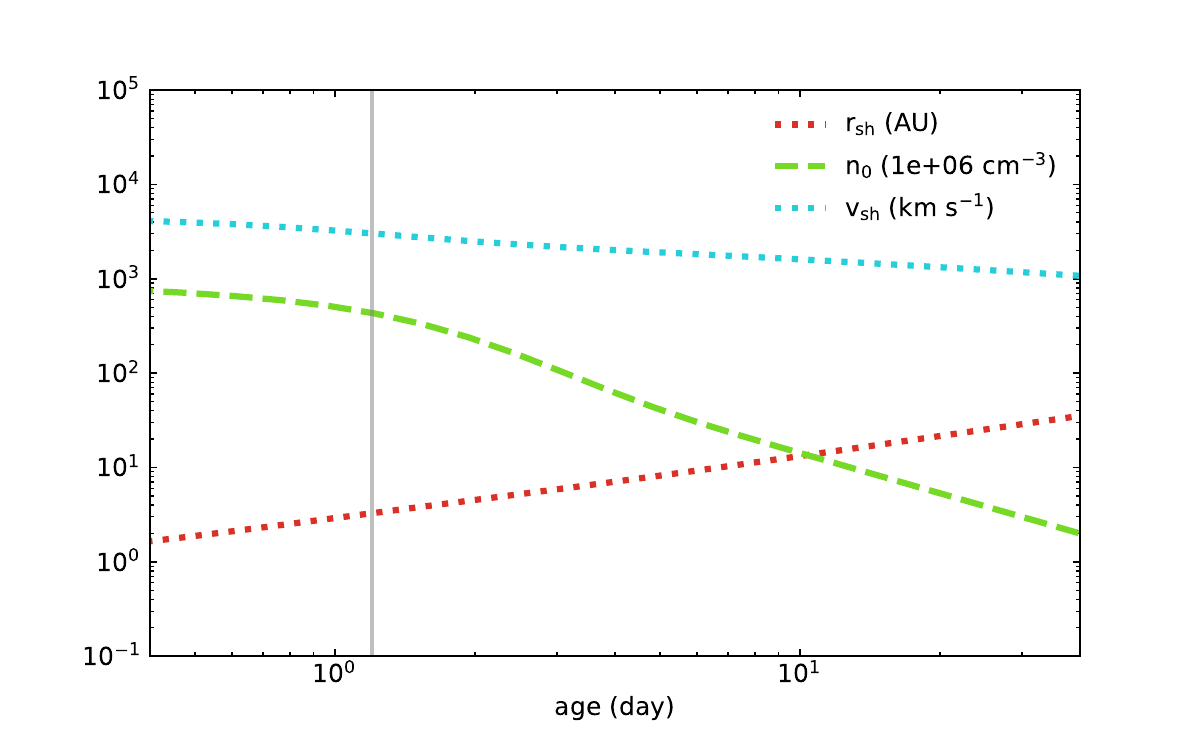}
    } 
    \\[-1em]
    \subfloat[\label{fig:lc}]{%
      \includegraphics[width=0.48\textwidth, clip=true,trim= 10 10 40 35]{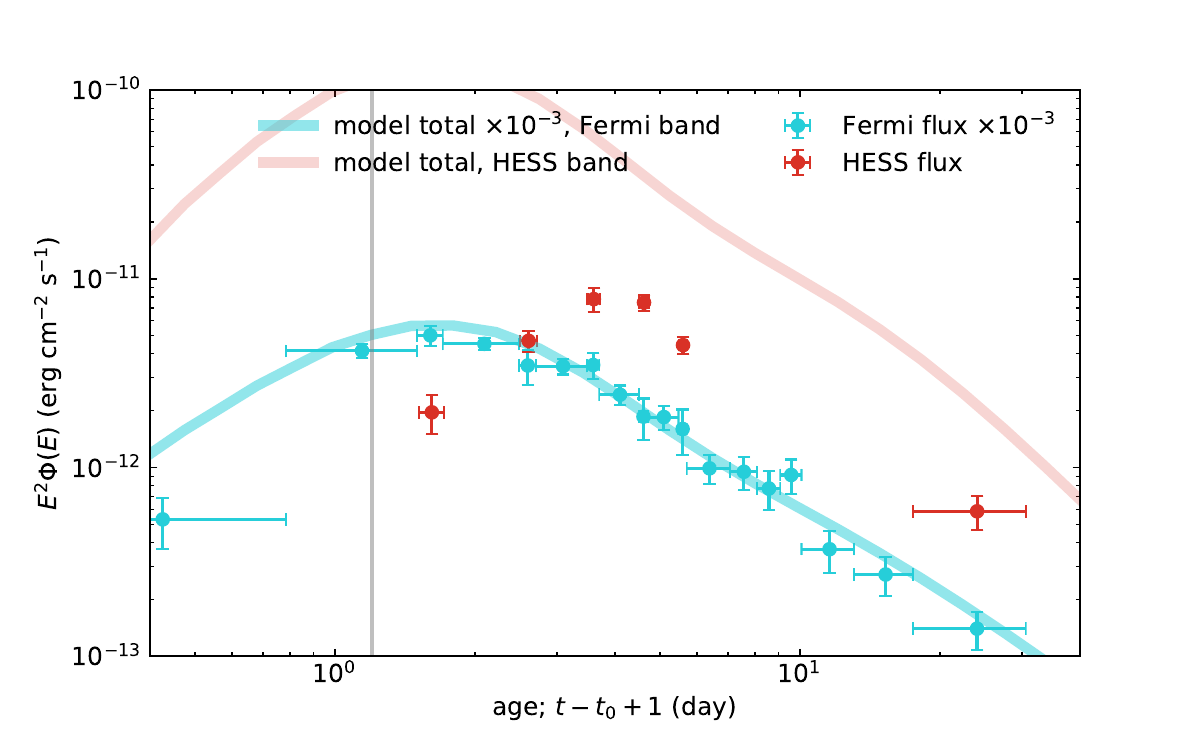}
    }

\caption{Hydrodynamic evolution (top) and light curve (bottom) for our single shock model that best fits the Fermi data \citep{HESS22}. 
Shock parameters are listed in Table \ref{tab:singleshock}. Gray lines indicate the approximate time of transition from the ejecta-dominated to the Sedov-Taylor stage.} For the overlaid observational data, $t_0$ corresponds to the peak of the optical light curve. 
The modeled light curves match the Fermi data quite well (bottom; blue line), but substantially overestimate the H.E.S.S. data (bottom; red line). 
This overestimation comes from the fact that the combined Fermi and H.E.S.S. data cannot be described as a single power law with an exponential cutoff, implying a more complex picture than a single, external shock.
\label{fig:singleshock}
\end{figure}

In this section we apply our model to RS Ophiuchi assuming its emission arises from a single, external shock. Broadly speaking, a viable model must reproduce three key features observed by Fermi, H.E.S.S., and MAGIC \citep{HESS22, MAGIC22, cheung+22}:

\begin{enumerate}
    \item An initial rise in both the GeV and TeV luminosities.
    \item An eventual decay in both the GeV and TeV luminosities that goes as $t^{-\alpha}$ where $\alpha \simeq 1.3-1.4$.
    \item A delay in the TeV luminosity peak with respect to the GeV luminosity peak of roughly 2-3 days.
\end{enumerate}

It is worth noting that, while H.E.S.S. observes a clear delay between GeV and TeV peaks, as well as a rise in the TeV emission at early times, MAGIC does not \citep{MAGIC22}. This discrepancy may be due to the fact that MAGIC began observing slightly later; we will therefore take the H.E.S.S. results at face value for the remainder of our analysis. Regardless, as we will show, the shape of the combined GeV-TeV spectrum alone requires the presence of multiple shock components.

The first two items on the above list can be reproduced with hadronic emission arising from a single external shock expanding into a medium of constant density that transitions to a red giant (RG) wind profile with density $\rho_0(r) \propto r^{-2}$ at large radii ($\gtrsim$ 3 AU).
Figure \ref{fig:singleshock} shows an example of shock evolution in such a density profile and the corresponding GeV and TeV light curves (more specifically, the light curves in the 0.06-500 GeV and 250-2500 GeV bands, consistent with the data displayed in \citealt{HESS22}).

Shock parameters, listed in Table \ref{tab:singleshock}, are chosen to be broadly consistent with observations and to provide a good fit to the Fermi (GeV) data. In particular, we adopt a RG wind velocity and mass loss rate that are comparable to (within a factor of $\lesssim 2$ of) those inferred observationally \citep[e.g., ][ finds a wind velocity of $33$ km s$^{-1}$ and a mass loss rate of $\sim 10^{-6} M_{\odot}$ yr$^{-1}$]{iijima09}. It is worth noting that we adopt a smaller ejecta mass than that inferred by \cite{pandey+22}. Since estimates of the ejecta mass are highly model-dependent, we prioritize a value that yields good agreement with Fermi observations (i.e., yields deceleration around the time of the Fermi luminosity peak). It is also worth noting that our ejecta mass is roughly consistent with those adopted elsewhere in the literature \citep[e.g, ][]{booth+16}.

A light curve peak that occurs $\sim$ 1 day after the shock is launched corresponds to a constant density region $\sim$ 3 AU, roughly twice the orbital distance of the RS Ophiuchi system \citep[e.g.,][]{booth+16}. The model GeV and TeV light curves predicted by this model are plotted in the bottom panel of Figure \ref{fig:singleshock}. 
The model produces a good fit to the observed GeV light curve. However, this model substantially overestimates the observed TeV emission, due to the fact that the outburst's observed spectrum is inconsistent with the theoretically-motivated power-law with an exponential cutoff.
A detailed discussion of this inconsistency and its implications can be found later in this section.

\begin{table}[ht]
    \centering
    \begin{tabular}{l l}
         Parameter                                  & Quantity (single shock) \\
         \hline
         \hline
         $M_{\rm ej}$ (ejecta mass)                 & $2\times10^{-7} M_{\odot}$ \\
         $v_{\rm sh, init}$ (initial $\vsh$)        & $4500$ km s$^{-1}$ \\
         $n_{\rm 0, init}$ (initial ambient density)  & $7\times10^{8}$ cm$^{-3}$\\
         $r_{\rm crit}$ (homogeneous region size) & $3$ AU \\
         $\dot{M}_{\rm wind}$ (RG mass loss rate)   & $5\times10^{-7} M_{\odot}$ yr$^{-1}$ \\ 
         $v_{\rm wind}$ (RG wind velocity)          & $30$ km s$^{-1}$ \\
         \hline \\
    \end{tabular}
    \caption{Model parameters used to fit the Fermi light curve, assuming a single, external shock. 
    The resulting hydrodynamic evolution and light curves can be found in Figure \ref{fig:singleshock}. 
    Note that a smoothing function is applied to the density profile, such that the density-related parameters shown here differ slightly from the ultimate profile that enters into our model. Without including a smoothing function, our model still reproduces key features of the GeV data, with the exception of a sharp jump in the light curve at the transition from the homogeneous region to the wind profile.}
    \label{tab:singleshock}
\end{table}

The luminosity rise, decay, and corresponding density profile requirements shown in Figure \ref{fig:singleshock} are best understood in terms of simple scaling relations. 
Namely, the luminosity of accelerated protons scales with the energy flux across the shock multiplied by the area of the shock surface $L_{\rm p} \propto \rho_0\vsh^3\rsh^2$.
For hadronic emission (i.e., pion decay), the $\gamma$-ray luminosity, $L_{\gamma}$, scales with the proton luminosity, the target density, and the shock age:  $L_{\gamma} \propto \rho_0^2\vsh^2\rsh^3$. 
Assuming a power-law scaling of the ambient density, $\rho_0 \propto \rsh^{-s}$, 
\begin{equation}
    L_{\gamma} \propto t^{3-2s},
\end{equation}
 during the ejecta-dominated stage, when the shock velocity is roughly constant and $\rsh \propto t$. 
Thus, a rise in luminosity at early times is only possible if $s<3/2$, suggesting that the material in the region closest to the white dwarf does not follow a wind profile ($s=2$).
Practically speaking, a flat profile ($s=0$) at small radii reproduces the observed luminosity rise quite well.  The actual CSM profile surrounding the white dwarf, as fed by the wind and Roche-lobe overflow of the giant, is expected to be complex \citep{martin+13,booth+16}.   

During the Sedov stage, $\rsh \propto t^{2/(5-s)}$ yielding,
\begin{equation}
 L_{\gamma} \propto t^{-2s/(5-s)}.   
\end{equation}
Thus, if the shock enters the Sedov stage while the ambient density is still relatively constant, $L_\gamma$ is roughly constant.
However, if the shock enters the Sedov stage around or after the ambient material has transitioned to a wind, $L_{\gamma} \propto t^{-4/3}$, consistent with the decays observed by both Fermi and H.E.S.S.

Together, these scaling relations suggest that RS Ophiuchi's outburst expanded first into a roughly homogeneous medium on scales comparable to the binary orbital separation, followed by the $\rho_0 \propto r^{-2}$ spherical red-giant wind on larger radial scales.
The fact that the observed GeV and TeV light curves rapidly transition from rising to power-law decay with $\alpha \simeq 0.3-0.4$ suggests that the onset of the Sedov stage roughly coincides with the start of the wind.
This coincidence allows us to place constraints on the extent and density of the homogeneous region.
Namely, the total mass in this region must be approximately equal to the ejecta mass.
This requirement informs the model shown in Figure \ref{fig:singleshock}. 

To avoid sharp jumps in our modeled light curve, we use a smoothing function to connect the uniform and wind-profile regions of our models.  In reality, the CSM distribution in this messy transition, from the ``disk-like" CSM near the white dwarf to the RG wind, is likely to be complex and not spherically symmetric \citep{booth+16}; our adopted smoothing thus also crudely accounts for variations in the shock evolution across different solid angle sectors.

\begin{figure}
    \centering
    \includegraphics[width=0.48\textwidth, clip=true,trim= 10 10 40 35]{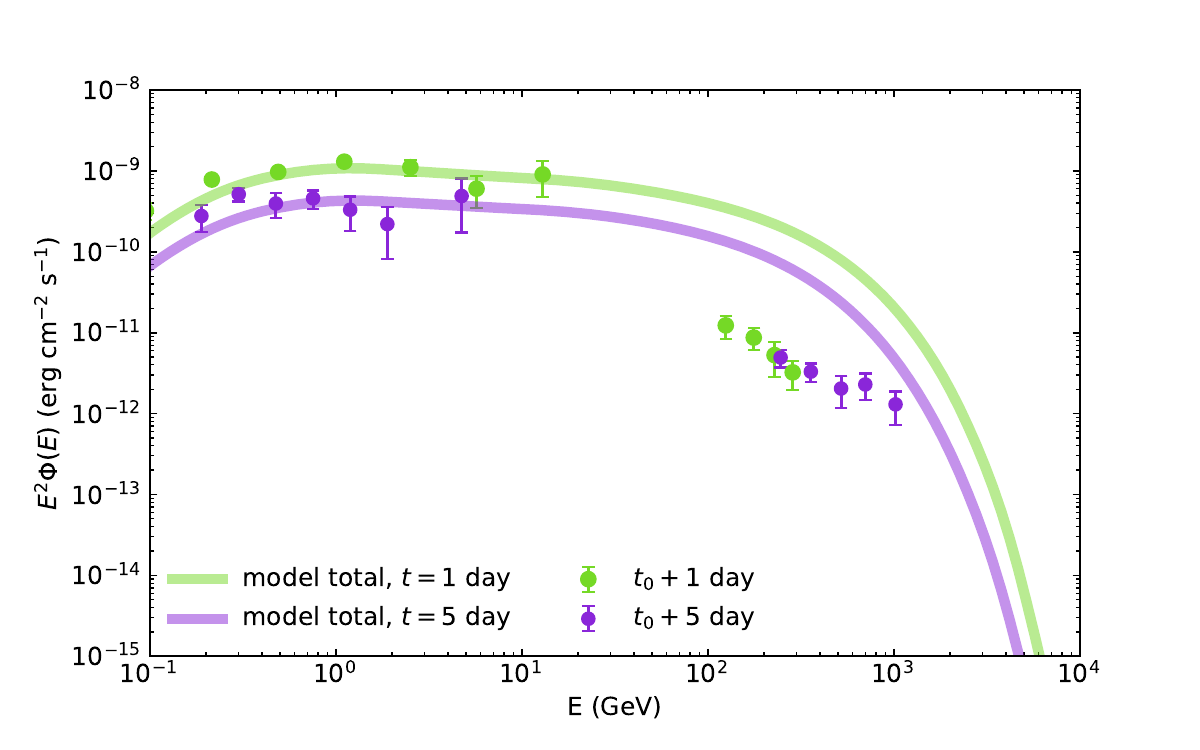}
    \caption{$\gamma$-ray spectra from our single-shock model with parameters described in Table \ref{tab:singleshock}. 
    The color scale of the lines and data points denotes the shock age and, in the case of the data points, the time after optical peak. 
    A single shock, which produces a power-law particle distribution with an exponential cutoff, cannot simultaneously describe the Fermi and H.E.S.S. data.}
    \label{fig:singleshock_spectra}
\end{figure}

While the single-shock model can describe the overall shapes of both the GeV and TeV light curves, it is inconsistent with observations in two key ways. 
First, and most obviously, it overestimates the very high energy (VHE) $\gamma$-ray flux by more than an order of magnitude. This arises from the fact that the combined Fermi and H.E.S.S. data are inconsistent with a power-law $\gamma$-ray spectrum with an exponential cutoff. 
In particular, the H.E.S.S. spectrum five days after the optical peak follows a steep power-law ($\propto E^{-q}$ where $q \gg 2$) or the very beginning (i.e., the low-energy, slowly falling portion) of an exponential cutoff.  However, the overall normalization of this spectrum falls well below the overall normalization of the Fermi spectrum on the same day.  As a result, theoretically-motivated models (i.e., power law spectra with exponential cutoffs) that fit the Fermi data will either overestimate the H.E.S.S. observations or, if a sufficiently small $\vsh$ is chosen to reduce the maximum energy, produce VHE spectra that is properly normalized but much steeper than that observed.  To illustrate this issue, Figure \ref{fig:singleshock_spectra} shows $\gamma$-ray spectra from our single shock model, which fit the Fermi data and have maximum energies in the needed range, but overestimate the VHE flux.

We note that MAGIC \citep{MAGIC22}, HESS \citep{HESS22}, and \cite{zheng+22} interpret the combined Fermi and VHE spectra as arising from a single shock. 
To fit these combined spectra, H.E.S.S. evokes a slow exponential cutoff ($\propto e^{-(E/E_{\rm max})^\beta}$), where $\beta < 1$. 
This modification results in a good fit to the data but is not theoretically motivated. Meanwhile MAGIC does fit their spectra with a theoretically motivated power-law with exponential cutoff. However, they achieve this fit by invoking arbitrary normalizations and maximum energies that do not evolve in a physical manner.

A second key tension regarding the single-shock model is that it cannot reproduce the delay between the GeV and TeV peaks. 
As our model demonstrates, a single shock yields a luminosity peak that occurs at approximately the same time for all energies.  However, as put forward in \cite{HESS22}, the VHE peak may be modulated by the maximum energy or, equivalently, by the finite acceleration time for TeV particles.

To illustrate why such a modulation cannot resolve the time delay issue, let us consider some simple scaling relations.
Assuming $E_{\rm max}$ is set by requiring that the acceleration time be approximately equal to the diffusion time, and that this time be less than or equal to the age of the system, we find that $E_{\rm max} \propto B_2\vsh^2 t$ for Bohm diffusion \citep{caprioli+14c}. 
Here, $B_2$ is the magnetic field behind the shock, $\propto \rho_0^{1/2} \vsh^{3/2}$ if the non-resonant streaming instability dominates magnetic field amplification.
Thus, we find $E_{\rm max} \propto \rho_0^{1/2}\vsh^{7/2} t$.
This framework for $E_{\rm max}$ is broadly equivalent to that in our model, in which the diffusion length of the highest-energy particles is a fixed fraction of the system size.

Alternatively, since the non-resonant instability is driven by escaping particles, one can assume, as in \cite{HESS22}, that the number of escaping particles with energy $E_{\rm max}$ is sufficient to drive the non-resonant instability and thereby inhibit particle propagation. 
This requirement yields $E_{\rm max} \propto \rho_0^{1/2}\vsh^2R_{\rm sh}$ \citep[see, e.g.,][for a detailed derivation]{bell+11}. 
In either case, the maximum energy would be predicted to increase prior to the GeV luminosity peak, when the shock velocity and$—$given arguments presented earlier in this section$—$the density, are roughly constant.  However, after the GeV luminosity peak, the shape of the light curve demands that the shock enter the Sedov stage.
During this stage, both of the assumptions outlined above give a maximum energy that decreases with time, regardless of whether the ambient density is constant or follows a wind profile. 
Thus, a rise in the maximum particle energy cannot account for the delayed VHE luminosity peak.

It is also worth noting that, to produce good agreement with the GeV data, we require the onset of the Sedov-Taylor phase occur at $t \simeq 1$ day (see the gray lines in Figure \ref{fig:singleshock}).
This result is inconsistent with the results of \cite{pandey+22} and \cite{cheung+22}, which find that the free expansion phase lasts until approximately day 4.
The fact that the TeV emission also does not peak until this time provides further evidence for multiple shock components.

Finally, we note that IC emission cannot resolve the issues mentioned here or, more to the point, contribute significantly to the $\gamma$-ray spectrum.  Cosmic ray (CR) electrons suffer strong synchrotron losses in the amplified fields calculated in our model ($\sim$ 1 G), severely reducing their ability to produce substantial VHE emission at any point during the shock's evolution.  More importantly, the large radiation fields expected near the forward shock ($\sim 1$ erg cm$^{-3}$, see the supplementary materials of \citealt{HESS22}) give an IC loss time $\simeq 15$ seconds for TeV electrons, less than their acceleration time.

\section{A Multi-Shock Scenario} \label{sec:results2}

\begin{figure}[ht]
  \centering
  \subfloat[\label{fig:evo_slow}]{%
      \includegraphics[width=0.48\textwidth, clip=true,trim= 35 30 40 35]{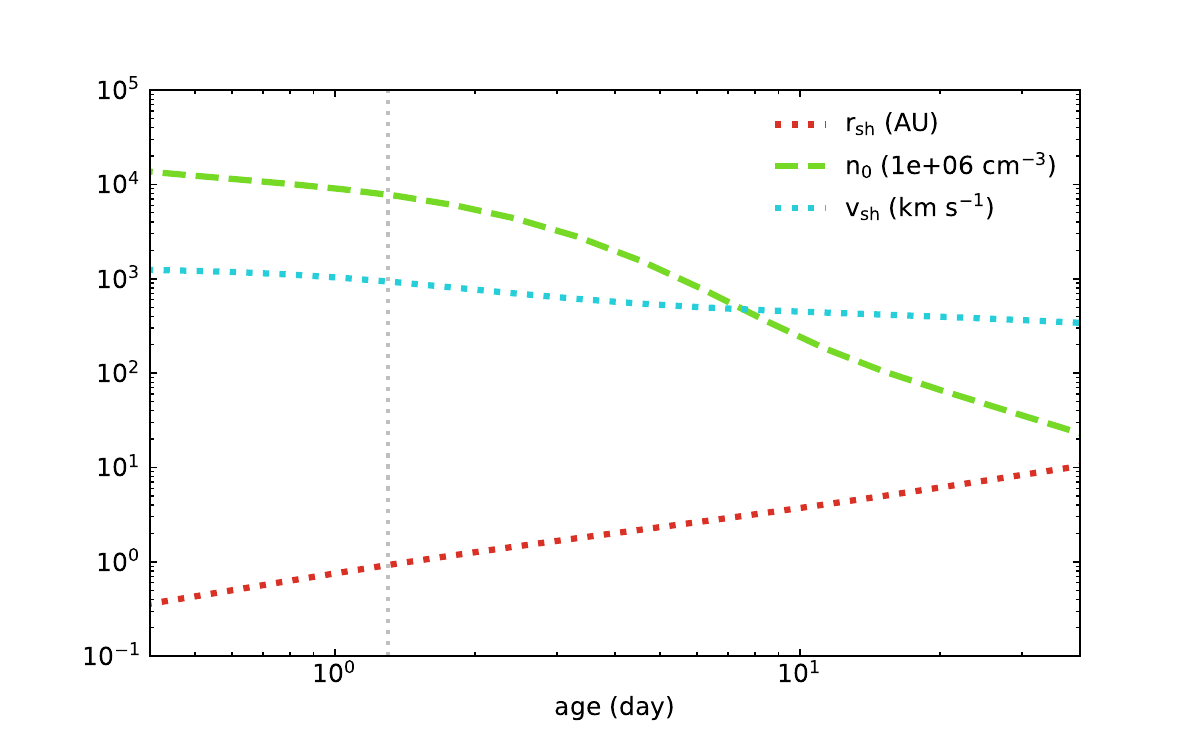}
    } 
    \\[-1em]
    \subfloat[\label{fig:evo_fast}]{%
      \includegraphics[width=0.48\textwidth, clip=true,trim= 35 10 40 35]{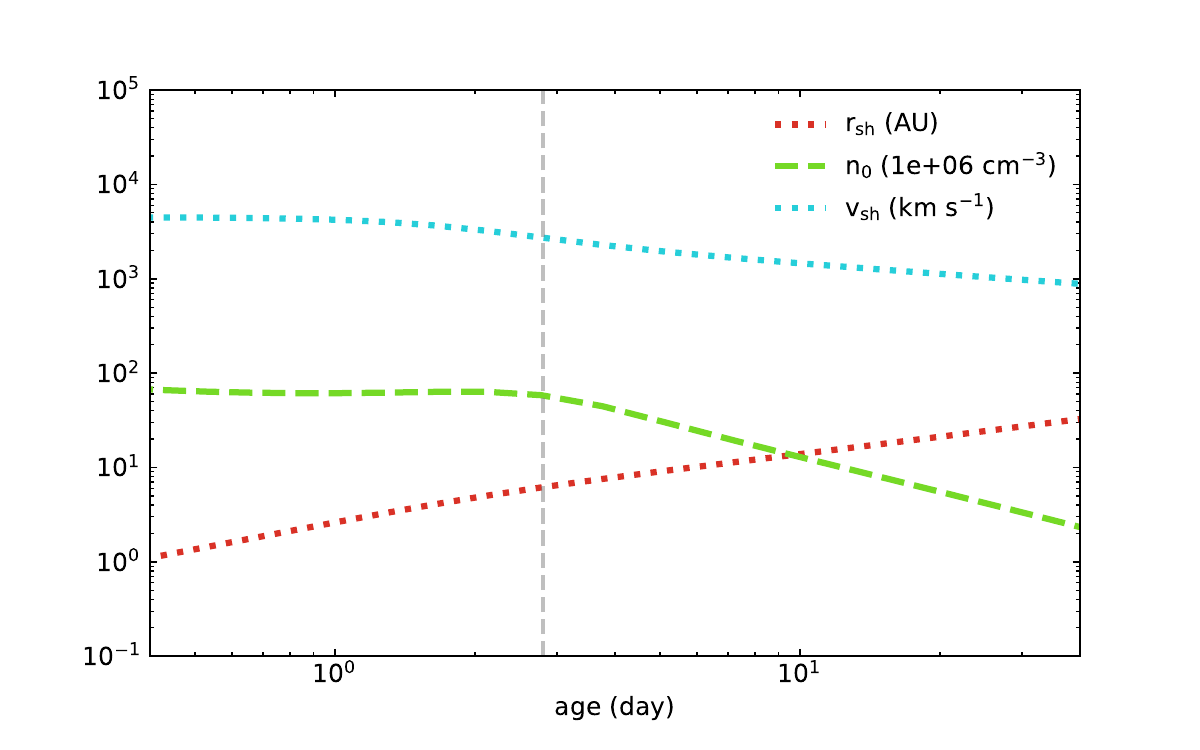}
    }

\caption{Hydrodynamic evolutions of the slow (top) and fast (bottom) components of our best-fit multi-shock model.
Gray lines indicate the approximate time of transition from the ejecta-dominated to the Sedov-Taylor stage.
Shock parameters are listed in Table \ref{tab:multishock}.}
\label{fig:multishock_evo}
\end{figure}

The previous section demonstrates that a single external shock cannot describe both the GeV and TeV emission from RS Ophiuchi's 2021 outburst, unless spectra and maximum energy scaling are chosen ad-hoc rather than calculated based on the DSA theory.
In this section, we instead explore a scenario involving two shocks which initially expand into different, roughly homogeneous media, but eventually probe the same RG wind on large scales.  Specifically, we aim to test whether multiple shocks, be they internal interactions between distinct ejecta components or a manifestation of a single ejecta running into an aspherical medium, could feasibly explain the observed $\gamma$-ray emission.  
Conversely, this exercise shows how the observed $\gamma$-ray emission places a constraint on the environment within and surrounding the nova system.

We shall focus our efforts only the simplest scenario--in terms of number of shock components--that reproduces the main features of the $\gamma$-ray observations. There may exist more complex scenarios that also yield a good fit to the GeV and TeV data.  Our goal here is simply to provide further evidence for a complex outburst involving multiple shock components.

\begin{figure*}[ht]
  \subfloat[\label{fig:multishock_lc}]{%
      \includegraphics[width=0.5\textwidth, clip=true,trim= 10 10 40 35]{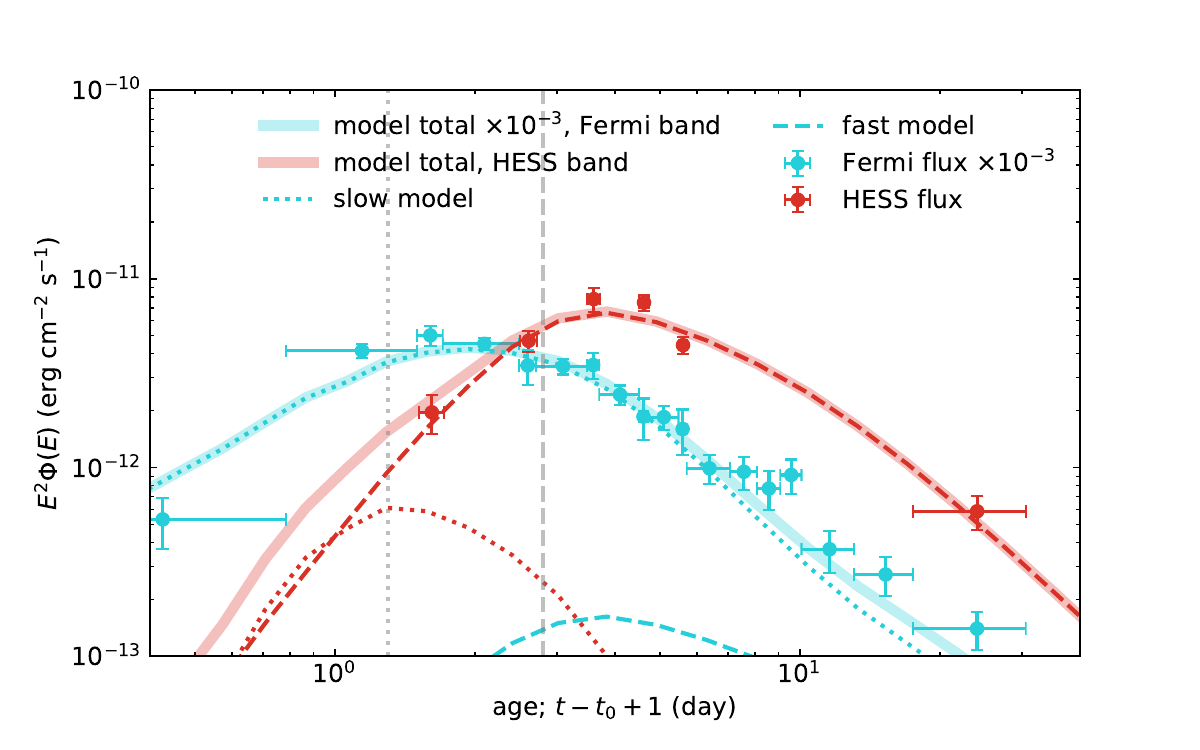}
    } 
    \subfloat[\label{fig:multishock_spec}]{%
      \includegraphics[width=0.475\textwidth, clip=true,trim= 35 10 40 35]{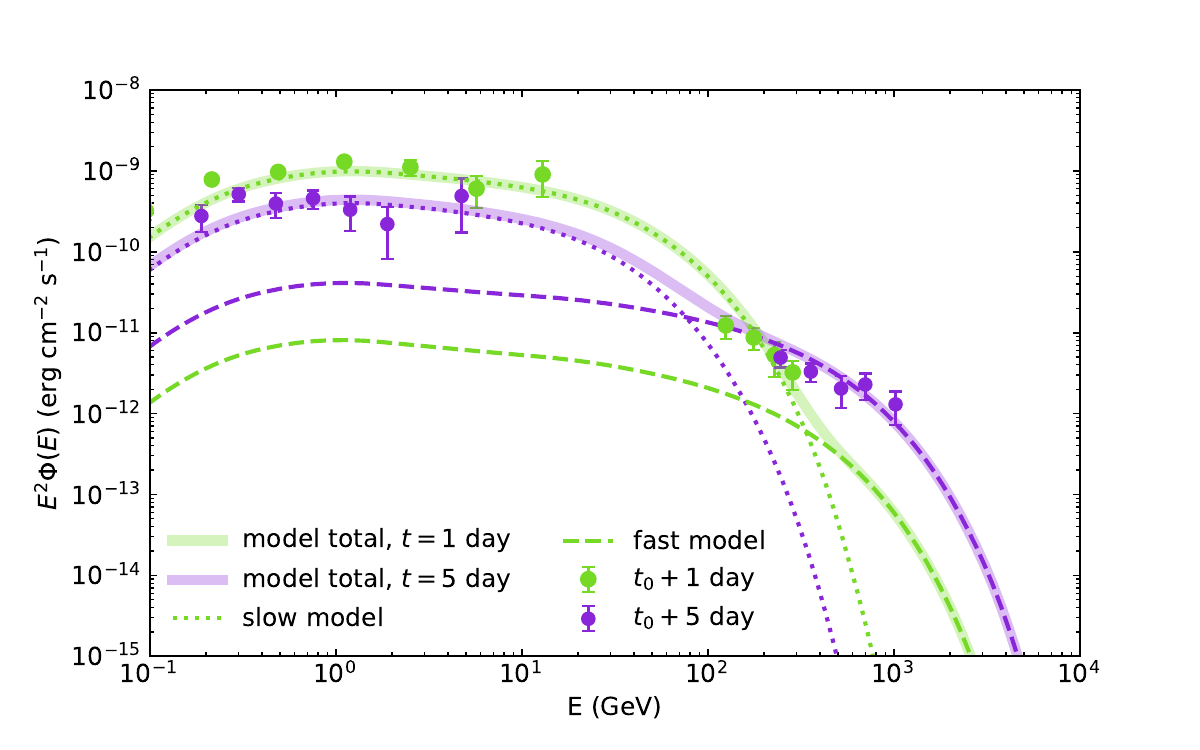}
    }

\caption{{\it Left:} Light curves from our two-shock model, displayed in the bands observed by Fermi (blue lines), and H.E.S.S. (red lines). The slow and fast components are shown with dotted and dashed lines, respectively, while the solid line gives their sum. Gray lines indicate the approximate time of transition from the ejecta-dominated to the Sedov-Taylor stage.}
{\it Right}: Modeled spectra after one and five days.  As in Figures \ref{fig:singleshock} and \ref{fig:singleshock_spectra}, all times are given in terms of shock age for the model and relative to $t_0$, the time of optical peak, for the observational data.  This two-shock model reproduces both the $\gamma$-ray light curves and spectra observed by Fermi and H.E.S.S. \\
\label{fig:multishock}
\end{figure*}

The simplest scenario that fits both the outburst's GeV and TeV light curves consists of two shocks: a slow, highly luminous component (i.e., a component that probes a relatively high density), and a fast, less luminous component (i.e., a component that probes a relatively low density).

The evolution of these components are shown in Figure \ref{fig:multishock_evo} and the corresponding model parameters are listed in Table \ref{tab:multishock}. We adopt the same RG wind parameters as in Section \ref{sec:results1}, which are consistent with the results of \cite{iijima09}.
In this picture, the slow component produces the bulk of the GeV emission along with the very steep TeV spectrum at early times, while the fast component produces the hardened TeV emission at later times.  This TeV hardening occurs because the fast component achieves both a higher maximum $\gamma$-ray energy and, since it sweeps up less mass at early times, a later luminosity peak. Note that the fast component of our multi-shock model has a Sedov-Taylor time of $t\simeq 3$ days, in better agreement with the deceleration timescale implied by X-ray observations \citep{pandey+22, cheung+22}.

The shock velocities we adopt are chosen to produce the best agreement with the observed $\gamma$-ray data. While these velocities are broadly consistent with optical data (specifically, $v_1$ and $v_2$ denoted in Section \ref{sec:spectroscopy}, they differ somewhat from those inferred from X-ray data \citep[][]{cheung+22}. This discrepancy may be the result of additional shock components, but may also be attributed to uncertainties arising from the conversion of X-ray temperatures to shock velocities \citep[see ][ for a detailed discussion]{orio+22}.

\begin{table}[ht]
    \centering
    \begin{tabular}{l l l}
         Parameter  & Slow Component    & Fast Component \\
         \hline
         \hline
         $M_{\rm ej}$   & $1\times10^{-7} M_{\odot}$    & $1\times10^{-7} M_{\odot}$ \\
         $v_{\rm sh, init}$ & $1300$ km s$^{-1}$     & $4500$ km s$^{-1}$ \\
         $n_{\rm 0, init}$   & $1.2\times10^{10}$ cm$^{-3}$ & $5.0\times10^{7}$  cm$^{-3}$\\
         $r_{\rm crit}$ & $1.0$ AU  & $6.0$ AU \\
         $\dot{M}_{\rm wind}$  & $5\times10^{-7} M_{\odot}$ yr$^{-1}$  & $5\times10^{-7} M_{\odot}$ yr$^{-1}$ \\ 
         $v_{\rm wind}$          & $30$ km s$^{-1}$ &   $30$ km s$^{-1}$\\
         \hline \\
    \end{tabular}
    \caption{Model parameters used to fit the Fermi and H.E.S.S. data, assuming the observed $\gamma$-ray emission arises from two shocks: a slow component that probes a relatively high ambient density, and a fast component that probes a relatively low ambient density. 
    After expanding through these different regions, both components probe the same RG wind.
    The resulting hydrodynamic evolution can be found in Figure \ref{fig:multishock_evo}. 
    Note that, as with the single shock model, a smoothing function is applied to the density profile, such that the density-related parameters shown here might differ slightly from the ultimate profile that enters into our model.}
    \label{tab:multishock}
\end{table}

The resulting light curves and spectra are shown in Figure \ref{fig:multishock}. 
This two-shock model yields good agreement with both the GeV and TeV observations.
In particular, because the fast component takes longer to reach its luminosity peak and has a higher $E_{\rm max}$ at that time, it reproduces the observed TeV delay.

For the sake of simplicity, we take the total $\gamma$-ray flux to be the sum of the fluxes from the two components.  A different set of model parameters might also fit the data if the relative weights of the two components are modified (i.e., if each component has a angular filling fraction $f \equiv \Delta \Omega/4\pi$ not equal to unity, where $\Delta \Omega$ is the solid angle subtended by the shock), or if the two shocks are launched at different times.  Broadly speaking, however, the two components must satisfy several conditions:

\begin{enumerate}
    \item \label{item:velocity} Shock velocities must differ by a factor of $\gtrsim 3$. This requirement comes from the need for sufficient stratification in $E_{\rm max}$ (i.e., the fast component must produce substantial TeV emission while the GeV component must not). If the two shock velocities are too similar, we revert to a scenario akin to a single-shock.
    \item \label{item:rcrit} Assuming the two shocks are launched simultaneously, the extent of the inner homogeneous region probed by the fast component, $r_{\rm crit}$, is $\gtrsim 6$ times larger than that of the slow component. As discussed in Section \ref{sec:results1}, the peak of a component's light curve roughly corresponds to when the shock starts to slow down and to probe the RG wind. Taking the shock age at the TeV peak to be roughly twice that at the GeV peak (consistent with observations), and the fast shock to have a velocity that is 3 times that of the slow shock, we obtain $r_{\rm crit, f}/r_{\rm crit, s} \simeq 6$, where subscripts f and s denote the fast and slow components, respectively.
    \item \label{item:rho0} Assuming equal contributions from the two components (i.e., each has the same filling factor, $f$), the uniform densities probed by the two components at small radii must differ by a factor of at least a few hundred. The GeV peak luminosity, produced by the slow component, is approximately a thousand times the TeV peak luminosity, produced primarily by the fast component. 
    With the modestly steep photon spectrum $dN/dE \propto E^{-2.3}$ returned by our self-consistent model \citep[see][for the physics behind the obtained spectral slope]{haggerty+20,caprioli+20,diesing+21}, we expect the GeV luminosity of the fast component to be roughly 10 times its TeV luminosity. Thus, at GeV energies, $L_{\gamma, \rm f}/L_{\gamma, \rm s} \sim 10^{-2}$, where these $L_{\gamma}$ are the maximum luminosities of each component. Since, at peak, $L_{\gamma} \propto \rho_0^2 \vsh^2 r_{\rm crit}^{3}$, our constraints on the ratios of $\vsh$ and $r_{\rm crit}$ give $\rho_{\rm 0, f}/\rho_{\rm 0, s} \simeq 2 \times 10^{-3}$. If we additionally introduce filling factors, $f_{\rm f}$ and $f_{\rm s}$ such that $f_{\rm f} = 1 - f_{\rm s}$ and $L_{\gamma} \propto f$ for each component (equivalent to the presence of polar and equatorial components of the shock), we can recast this condition as $\rho_{\rm 0, f}/\rho_{\rm 0, s} \simeq 2 \times 10^{-3}(1/f_{\rm f}-1)^{1/2}$. Note that the model parameters described in Table \ref{tab:multishock} differ slightly from this estimate, primarily due to the fact that our model is multi-zone and therefore includes contributions to $L_{\gamma}$ from multiple epochs of shock evolution.
    \item \label{item:Mej} In the case that $f_f = f_s = 0.5$, the two components must have comparable ejecta mass (i.e., within a factor of a few). Recall that, to reproduce the shape of the peak, both components must enter the Sedov stage when they reach the extent of their homogeneous region, $r_{\rm crit}$. Thus, the ejecta mass, $M_{\rm ej}$ must be equal to the mass contained in the homogeneous region, $\propto f\rho_0r_{\rm crit}$. For equal filling factors, the constraints enumerated above yield $M_{\rm ej, f} \simeq  0.4 M_{\rm ej, s}$. Otherwise, we have $M_{\rm ej, f}/M_{\rm ej, s} \simeq 0.4 \sqrt{f_{\rm f}/(1-f_{\rm f})}$.
    
\end{enumerate}

The above arguments make clear that a degeneracy exists between the relative ejecta masses of the fast and slow components and their relative angular filling factors. A similar degeneracy also exists between these filling factors and the relative ambient densities.  Nevertheless, the main conclusion of this exercise still stands: while a single shock cannot reproduce the $\gamma$-ray observations, scenarios with multiple shocks (or multiple shock components) are capable of doing so.

\section{Conclusion}
We have modeled the $\gamma$-ray emission of RS Ophiuchi's recent outburst using a semi-analytic model of particle acceleration that self-consistently accounts for magnetic field amplification as well as the back-reaction of nonthermal particles.  We demonstrate the properties of the observed $\gamma$-ray emission is not consistent with a single, external shock. This inconsistency arises from the facts that: a) the finite acceleration time of TeV particles is too short to explain the delay between the GeV and TeV peaks as observed by Fermi-LAT and H.E.S.S.; and (b) the combined GeV-TeV spectra are inconsistent with theoretically-motivated emission models for a single shock, in particular a power-law with an exponential cutoff.

On the other hand, we find that the observed $\gamma$-ray emission is naturally reproduced in a scenarios involving multiple shocks.  In particular, both the spectra and GeV/TeV light curves are consistent with the combined emission from two shocks: one with a low initial velocity expanding into a dense ambient medium, and one with a fast initial velocity expanding into a comparatively rarefied medium.  Different combinations of shocks with non-equal filling factors may also be able to reproduce the observations, as could scenarios with three or more shocks. The key takeaway, then, is that RS-Ophiuchi's recent outburst must be more complicated than the single-shock scenario presented in the literature. 

The presence of multiple ejecta components in RS Ophiuchi (and, hence, multiple shocks) is also supported by time-dependent optical spectroscopy, which reveals as many as three distinct velocity components (Sec.~\ref{sec:spectroscopy}).  Multiple shocks are also supported by X-ray observations \citep{Page+22, orio+22}; \citet{orio+22} find evidence for at least two components of shocked plasma, at temperatures $kT \approx 0.75$ keV and $\approx 3$ keV, respectively, which could in principle be attributed to the ``slow'' and ``fast'' shocks in our two-shock scenario.  However, the fact that the peak of the hard X-ray light curve is delayed relative to that of either the GeV or TeV $\gamma$-ray emission, is challenging to understand and may point to the presence of yet additional shocks. Thermal X-rays from the shocks which dominate the $\gamma$-ray emission be absent at early times if they are either absorbed by a large neutral gas column ahead of the shock, or intrinsically suppressed in the multi-phase post-shock region (e.g., \citealt{Steinberg&Metzger18,Nelson+19}).

The ejecta from classical novae are also characterized by multiple velocity components (e.g., \citealt{Aydi+20b}), with the slower ejecta concentrated in the equatorial plane of the binary and the faster component likely representing a more spherical wind from the white dwarf which then expands more freely into the polar directions (e.g., \citealt{chomiuk+14}).  If embedded novae like RS Ophiuchi generate ejecta with a similar geometry, then multiple ``slow'' and ``fast'' shocks would be generated as these outflows collide with themselves (as in classical novae) and with the surrounding circumbinary environment, which on radial scales interior to the binary separation is also highly aspherical, being significantly denser in the binary equatorial plane than along the polar directions (e.g., \citealt{booth+16}).  If this interpretation is correct, then one lesson to be gleaned from RS Ophiuchi is that$-$despite vastly different circumbinary environments$-$the ejecta properties of embedded/symbiotic novae may be similar to classical novae.  This in turn might post a challenge to models in which the binary companion (whose orbital separation from the white dwarf is significantly smaller in classical novae than in symbiotic/embedded novae like RS Ophiuchi) plays a dominant role in shaping the equatorial ejecta in classical novae (e.g., \citealt{MacDonald80,Livio+90}). 

Given that symbiotic novae are often treated as rapidly-evolving analogs for supernova remnants \citep[e.g.,][]{martin+13}, the complex behavior of RS Ophiuchi revealed in this work is also an important reminder that novae are fundamentally different systems with their own unique properties. That being said, some of this behavior may also be relevant for young supernovae, particularly those expanding into nonuniform media \cite[e.g., ][]{thomas+22}.
\label{sec:conclusion}

\acknowledgements
R.D., D.C., and S.G. were partially supported by NASA (grants NNX17AG30G, 80NSSC18K1218, and 80NSSC18K1726) and the NSF (grants AST-1714658, AST-1909778, PHY-1748958, and PHY-2010240).  B.D.M. was supported in part by NASA (grant 80NSSC22K0807). E.A. acknowledges support by NASA through the NASA Hubble Fellowship grant HST-HF2-51501.001-A awarded by the Space Telescope Science Institute, which is operated by the Association of Universities for Research in Astronomy, Inc., for NASA, under contract NAS5-26555. L.C. and E.A. are grateful for support from NSF grant AST-1751874 and NASA grant 80NSSC20K1535. I.V. acknowledges support by the ETAg grant PRG1006 and by EU through the ERDF CoE grant TK133.

\bibliographystyle{aasjournal}

\section{Supplementary plots}
\label{appA}
In this Appendix we present supplementary plots.

\begin{figure*}
\begin{center}
  \includegraphics[width=0.9\textwidth]{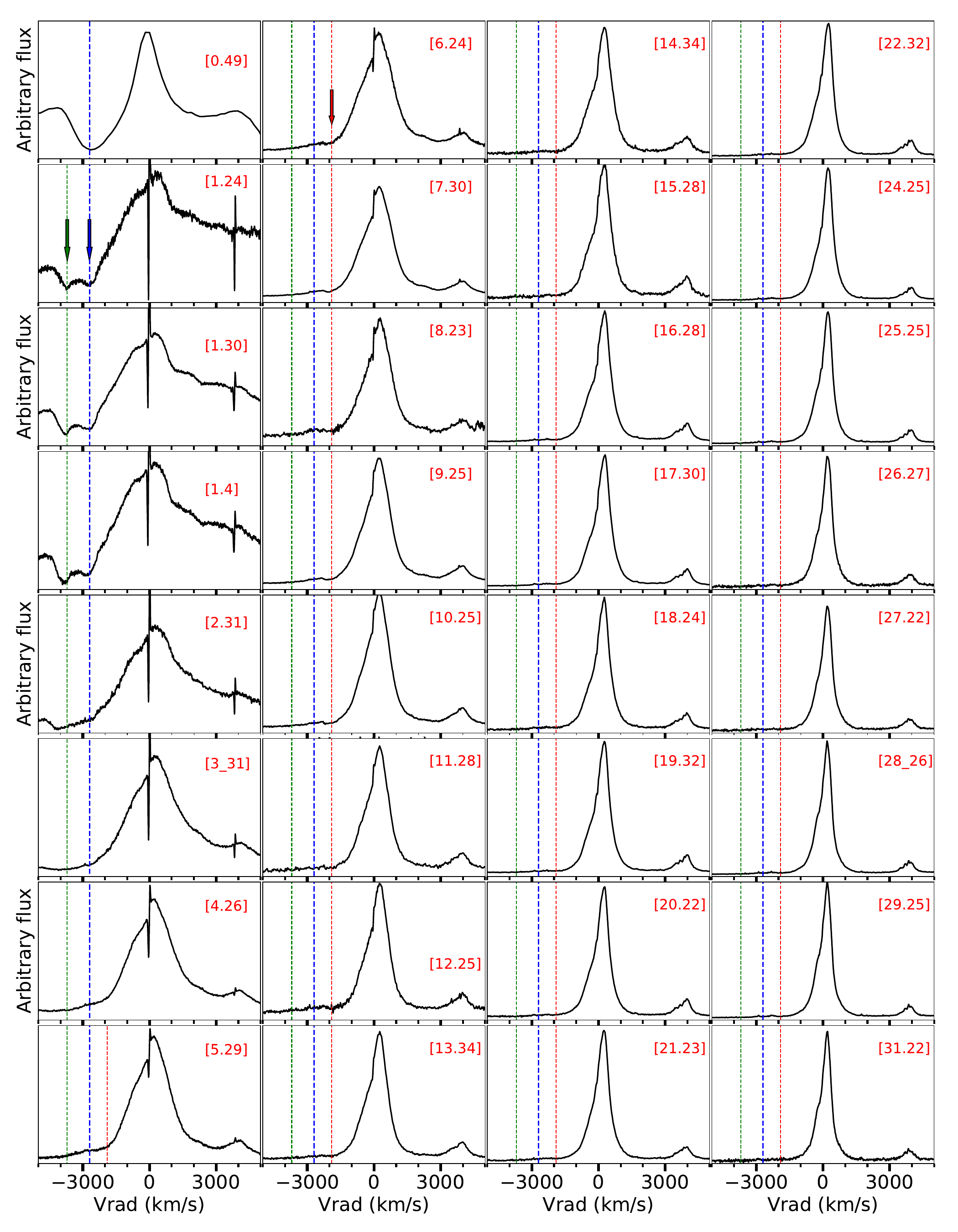}
\caption{The line profile evolution of H$\beta$ throughout the first 30 days of the eruption of RS~Oph. The numbers between brackets are days after $t_0$. The blue, green, and red dashed lines represent the velocities $v_1 = -2700$\,km\,s$^{-1}$, $v_2 = -3700$\,km\,s$^{-1}$, $v_3 = -1900$\,km\,s$^{-1}$, respectively. For clarity we also use arrows with the same corresponding colors, pointing to the line profiles. A heliocentric correction is applied to the radial velocities.} 
\label{Fig:Hbeta_complete}
\end{center}
\end{figure*}

\begin{figure*}
\begin{center}
  \includegraphics[width=0.9\textwidth]{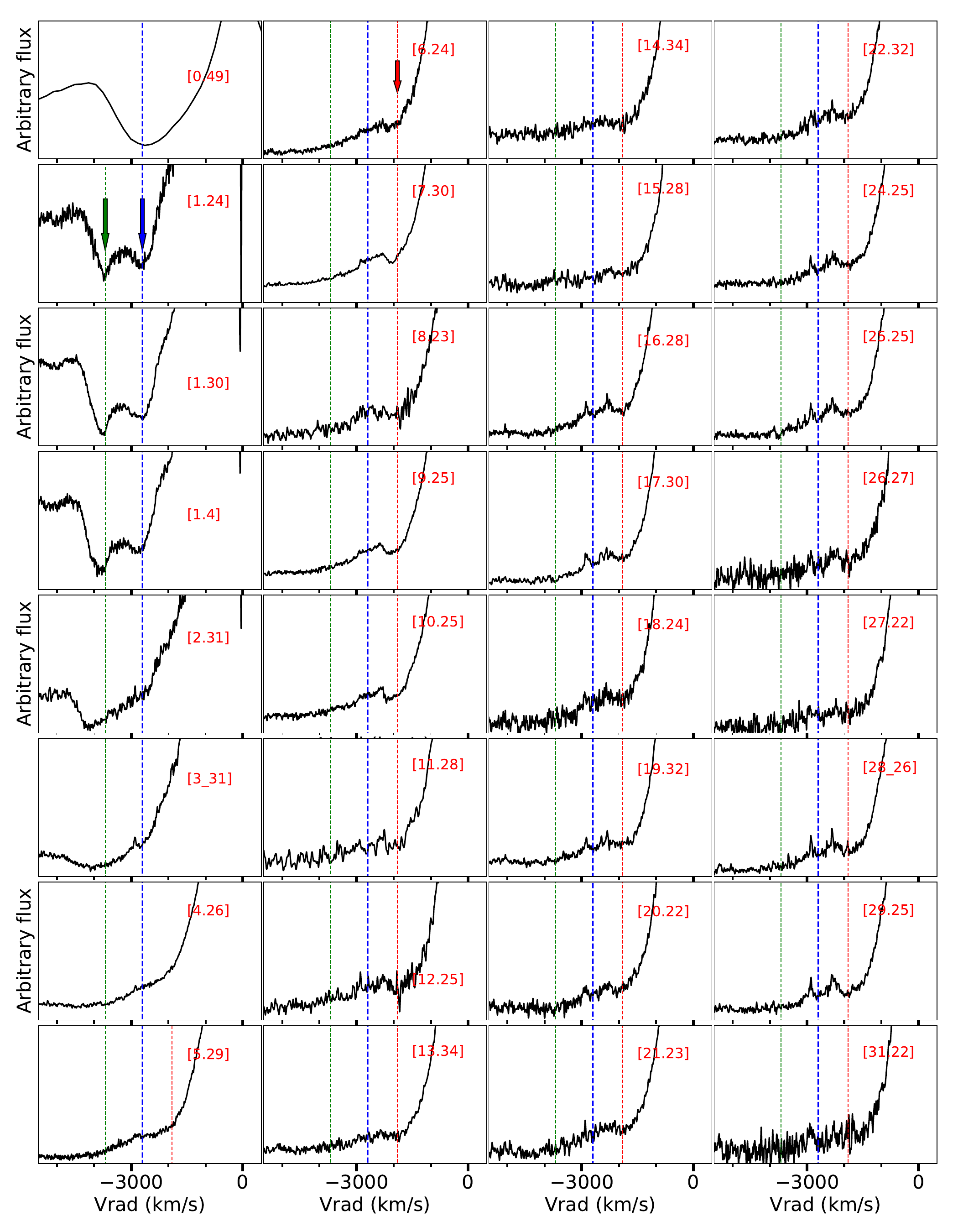}
\caption{Same as Figure~\ref{Fig:Hbeta_complete} but zooming in on the P Cygni absorption components.} 
\label{Fig:Hbeta_abs}
\end{center}
\end{figure*}

\begin{figure*}
\begin{center}
  \includegraphics[width=0.9\textwidth]{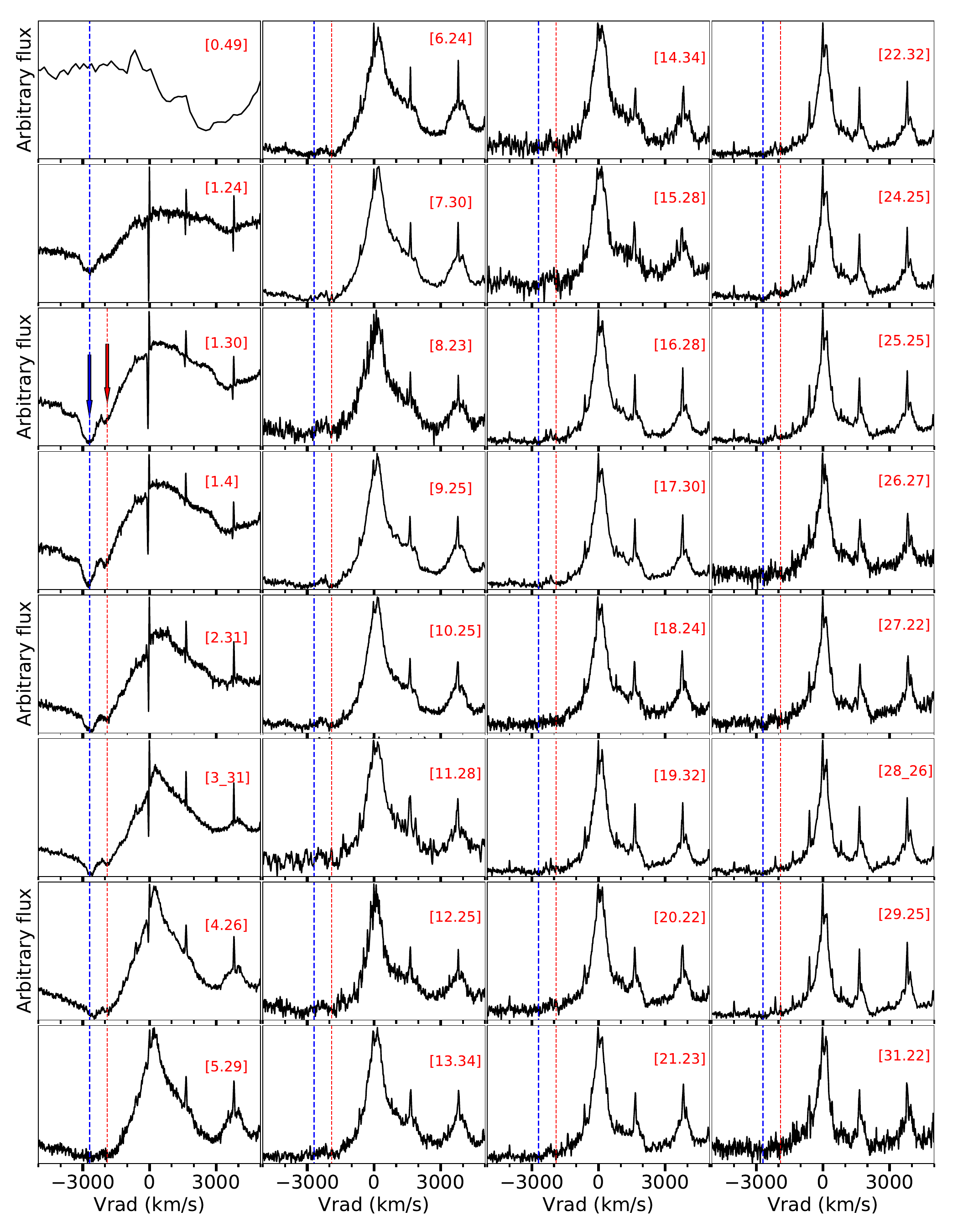}
\caption{The line profile evolution of Fe II 5018\,$\mathrm{\AA}$ throughout the first 30 days of the eruption of RS~Oph. The numbers between brackets are days after $t_0$. The blue, and red dashed lines represent the velocities $v_1 = -2700$\,km\,s$^{-1}$ and $v_3 = -1900$\,km\,s$^{-1}$, respectively. For clarity we also use arrows with the same corresponding colors, pointing to the line profiles. A heliocentric correction is applied to the radial velocities.} 
\label{Fig:FeII_complete}
\end{center}
\end{figure*}

\begin{figure*}
\begin{center}
  \includegraphics[width=0.9\textwidth]{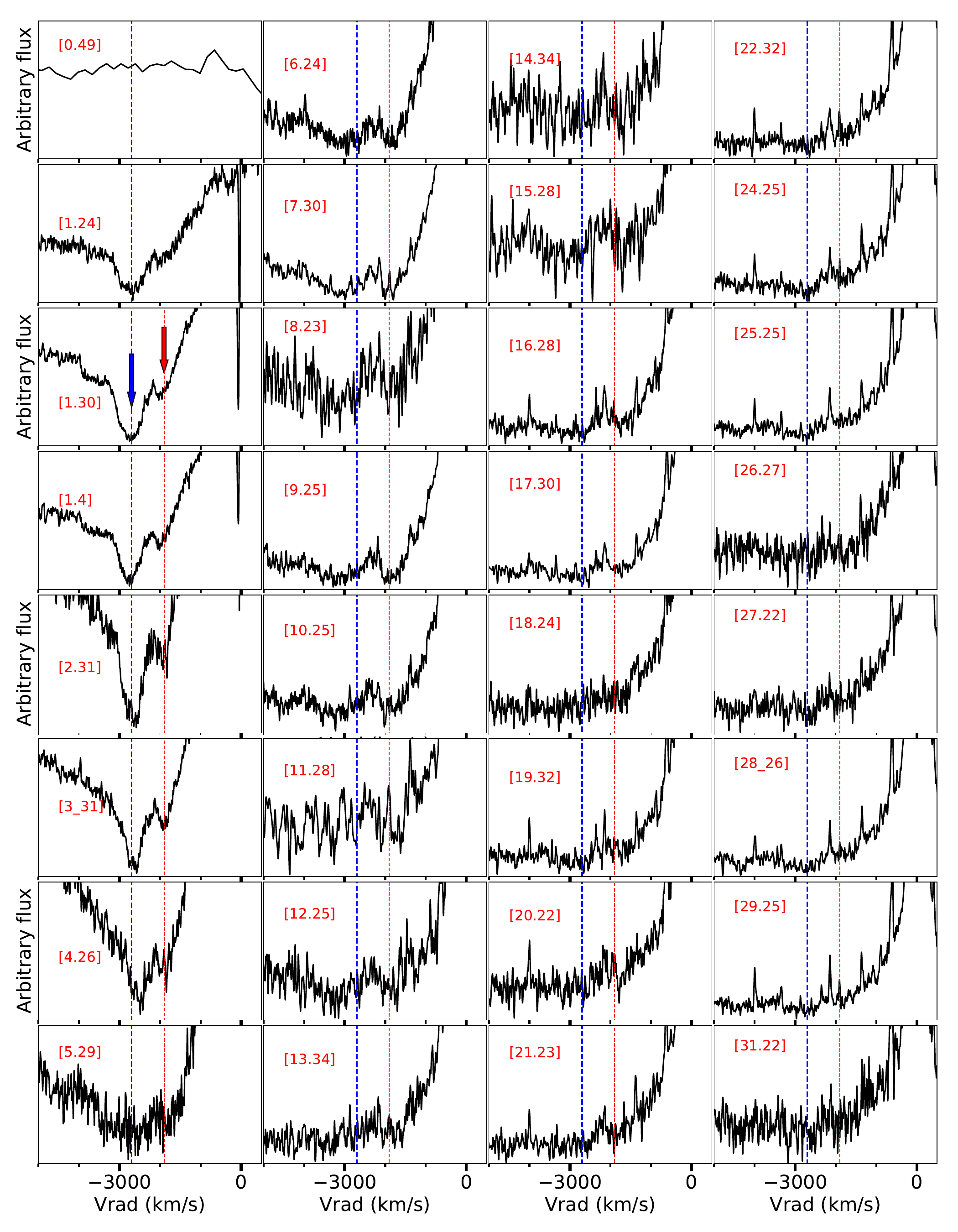}
\caption{Same as Figure~\ref{Fig:FeII_complete} but zooming in on the P Cygni absorption components.} 
\label{Fig:FeII_abs}
\end{center}
\end{figure*}

\end{document}